\documentclass[twocolumn,showpacs,
pra,aps]{revtex4}
\usepackage{array}
\usepackage{amsmath}
\usepackage{graphicx}
\usepackage{graphicx}
\usepackage{amsmath}
\usepackage{latexsym}
\usepackage{amsfonts}
\usepackage{amssymb}
\usepackage{array}
\usepackage{epsfig}

\newcommand{\Tr}{\mathrm{Tr}}
\newcommand{\diag}{\mathrm{diag}}

\begin{document}

\title{
 Generating ``squeezed'' superpositions of coherent states using photon addition and subtraction
 }

\author{P. Marek$^{1,3}$, H. Jeong$^2$, and M. S. Kim$^1$}
\affiliation{
$^1$School of Mathematics and Physics, The Queen's University,
Belfast BT7 1NN, United Kingdom\\
$^2$Center for Subwavelength Optics and Department of Physics and Astronomy, Seoul National
University, Seoul, 151-742, South Korea \\
$^3$Department of Optics, Palack\'{y} University, 17. listopadu 50, 77200 Olomouc, Czech Republic}
\date{\today}

\begin{abstract}
We study how photon addition and subtraction can be used to 
generate squeezed superpositions of coherent states in free-traveling fields (SSCSs) 
with high fidelities and large amplitudes.
It is shown that
an arbitrary $N$-photon subtraction
results in the generation of a SSCS with nearly the perfect fidelity ($F>0.999$) regardless
of the number of photons subtracted. In this case, the amplitude of the SSCS
increases as the number of the subtracted photons gets larger.
For example, two-photon subtraction from a squeezed vacuum state of 6.1dB
can generate a SSCS of $\alpha=1.26$, while in the case of
the four-photon subtraction a SSCS of a larger amplitude $\alpha=1.65$
is obtained under the same condition.
When a photon is subtracted from a squeezed vacuum state and another photon is added subsequently,
a SSCS with a lower fidelity ($F\approx 0.96$) yet higher amplitude ($\alpha\approx2$)
can be generated. We analyze some experimental imperfections including inefficiency of the detector
used for the photon subtraction.
\end{abstract}


\maketitle

\section{Introduction}

The development of the quantum theory of light has deepened our
understanding of nonclassical properties of optical fields.
Recently, superpositions of coherent states in free-traveling
fields (SCSs) \cite{Yurke,WScat} have attracted special attention
due to their remarkable usefulness. When their amplitudes are
large, the SCSs show typical properties of macroscopic quantum
superpositions, and because of this, they are often called
``Schr\"odinger cat states" recalling the famous cat paradox
\cite{Schr}. The SCSs enable one to perform many interesting
studies for fundamental tests of quantum theory
\cite{Derek,jeongsonkim,Magda}. Furthermore, it has been found
that SCSs are useful for various applications in quantum
information processing
\cite{Enk01,JKL01,Jeong02,Ralph03,WeakForce,puri}. The power of
this approach lies in the fact that all the four Bell states can
be discriminated in a deterministic way only using
a beam splitter and photon counting \cite{JKL01,puri}, which
is obviously not the case for the single-photon based approach.

In spite of the manifold usefulness of the SCSs, until recently,
the generation of free-travelling SCSs has been known to be
difficult. There have been schemes to generate such SCSs using
strong nonlinear interactions \cite{Yurke,gerry-added} or photon number
resolving detectors \cite{Dakna,Dakna2}, which are not feasible
using current technology. Recently, more realistic schemes have
been suggested by several authors
\cite{Lund04,jc1,jc2,jc3,jc4,KM}.
For example, a scheme  using weak Kerr nonlinearities and
simple optical elements was suggested \cite{jc2}
based on a previous proposal where strong
Kerr nonlinearities are required \cite{gerry-added}.
As another example, a simple observation was made that
SCSs with small amplitudes, such as $\alpha<1.2$, are well
approximated by squeezed single photons \cite{Lund04}. It was also
pointed out that squeezed single photons can be obtained by
subtracting or adding one photon from pure squeezed vacuums
\cite{JLR05}. Meanwhile, single-photon-subtracted squeezed states,
which are close to the SCSs with small amplitudes
($\alpha\lesssim 1$),
have been generated by several experimental groups
\cite{kt1,kt2,kt3,kt4} and theoretical analysis has been performed
\cite{SS06,Kim05,OP}. 
Recently, squeezed SCSs (SSCSs) were generated and detected
\cite{Cat07}, where the size of the states ($\alpha=1.6$) was
reasonably large for fundamental tests of quantum theory and
quantum information processing, for which the states are suited
despite their squeezing \cite{suitability}.
A scheme using time separated two-photon subtraction was suggested \cite{Sasaki1} and
experimentally demonstrated \cite{Sasaki2} to generate SCSs of large amplitudes.
Despite all the recent progress, however, the fidelities of the generated states are
yet to be improved for practical quantum information processing.

The directions of the development
for the generation of SCSs are twofold.
First, one needs to generate SCSs with larger amplitudes ($\alpha\gtrsim 2$)
for macroscopic tests of quantum theory.
Second, for quantum information processing, it is important to generate
SCSs with higher fidelity $F>0.99$ while $\alpha\approx 1.6$ is an appropriate value
\cite{Lund07}.
The SSCSs, generated in a recent experiment \cite{Cat07}, are simply a squeezed version of the SCSs. 
Interestingly, the direction of the squeezing 
in Ref.~\cite{Cat07} makes the SSCSs more robust against decoherence than the regular 
SCSs \cite{Serafini}.
The SSCSs can be useful in some protocols as they are \cite{suitability},
and if required, it may be possible to unsqueeze them by means
of the squeezing transformation \cite{jc3,jc4,squeez1}.

In this paper, we are interested in finding methods to
generate SSCSs with larger amplitudes and higher fidelities
using photon subtraction and addition.
Here, we show that the two photon subtraction enables one to produce
the SSCS with very high fidelity, and the combination of the subtraction
and addition can return the SSCS with
a lower fidelity yet higher amplitude.

We also find that consecutive applications of photon subtraction
(or subtracting a well defined number of photons) from a squeezed vacuum state
result in the generation
of a SSCS with nearly the perfect fidelity regardless
of the number of photons subtracted.
The amplitude of the SSCS
increases as the number of the subtracted photons gets larger.
This paper is organized as follows. In Sec.~II,
we investigate combinations of the ideal single photon addition and subtraction.
In Sec.~III, some experimental imperfections are analyzed for the realization of
the states discussed in Sec.~II.
In Sec.~IV, we numerically show that an arbitrary $N$ photon subtraction,
regardless of $N$, can be used to generate a SSCS with an
extremely high fidelity as $F>0.999$. We conclude with final remarks in Sec.~V.

\section{Photon addition to and subtraction from squeezed vacuum}

\subsection{Combinations of photon addition and subtraction}

The SCSs are defined as
\begin{equation}
\label{CSSdefine}
|{\rm SCS}_\varphi\rangle=
{\cal N}_\varphi(|\alpha\rangle+e^{i\varphi}|-\alpha\rangle),
\end{equation}
where ${\cal N}_\varphi$ is a normalization factor,
$|\pm\alpha\rangle$ is a coherent state of amplitude $\pm\alpha$,
and $\varphi$ is a real local phase factor.
The SCSs  such as
$|{\rm SCS}_\pm(\alpha)\rangle={\cal N}_{\pm}(|\alpha\rangle\pm|-\alpha\rangle)$
are called even and odd SCSs respectively because the even (odd) SCS
always contains an even (odd) number of photons.
The squeezed vacuum state, $|{\rm S}(r)\rangle$,
with the squeezing parameter $r$ can be obtained by applying the squeezing operator,
\begin{equation}
\hat{S}(r) = e^{\frac{r}{2}(\hat{a}^2 - \hat{a}^{\dagger 2})},
\end{equation}
where $\hat{a}$ ($\hat{a}^\dagger$) is the bosonic annihilation (creation)
operator, to the vacuum state.
In the number state basis, it can be represented as
\begin{equation}
|{\rm S}(r)\rangle =\sqrt{{\rm sech} r}
\sum_{k=0}^{\infty} \frac{\sqrt{(2k)!}}{k!}
\big[-\frac{1}{2}\tanh r\big]^k|2k\rangle,
\label{eq:sq}
\end{equation}
where
$r$ was assumed to be real.
As seen in Eq.~(\ref{eq:sq}),
the squeezed vacuum state contains only even number of photons.
A squeezed single photon, known as a good approximation
of a small SCS, can be obtained by adding a photon to
a squeezed vacuum as
\begin{equation}
\hat a^\dagger \hat{S}(r)|0\rangle=\cosh r \hat{S}(r)|1\rangle,
\label{addition}
\end{equation}
where the right hand side is unnormalized due to the characteristics of
the creation operator $a^\dagger$.
We note that Eq.~(\ref{addition}) can easily be shown using 
following unitary transformations \cite{BRbook}: 
\begin{equation}
\begin{aligned}
&\hat{S}^{\dag}(r)\hat{a}\hat{S}(r) = \hat{a} \cosh r -
\hat{a}^{\dag}\sinh r,\\
&\hat{S}^{\dag}(r)\hat{a}^\dagger\hat{S}(r) = \hat{a}^\dagger \cosh r -
\hat{a}\sinh r.
\label{iden}
\end{aligned}
\end{equation}
It is also known that the squeezed single photon can also be
obtained by subtracting a photon from a squeezed vacuum as
\begin{equation}
\hat a \hat{S}(r)|0\rangle=-\sinh r \hat{S}(r)|1\rangle.
\label{subtraction}
\end{equation}
This may cause us to conjecture that when the photon addition and subtraction are applied
successively to the squeezed vacuum, an approximate even SCS may be generated.
We can consider four immediate cases, namely, addition and subtraction ($ \hat a\hat a^\dagger$),
subtraction and addition ($\hat a^\dagger\hat a $),
successive additions ($(\hat a^\dagger)^2$),
and successive subtractions (${\hat a}^2$).
From Eqs.~(\ref{addition}) and (\ref{subtraction}),
it is straigtforward to see that $(\hat a^\dagger)^2$ will result in the same
state produced using $\hat a^\dagger\hat a $. It is also easy to see that
$\hat a\hat a^\dagger$ will cause the same effect with ${\hat a}^2$.
Therefore, we shall consider only two cases, $\hat a^\dagger\hat a $
and  ${\hat a}^2$, among the four based on the fact that the photon subtraction is
relatively easier to perform than the photon addition.

Now suppose an {\it ideal} situation that a photon is subtracted from the squeezed vacuum state and
then another photon is subsequently added.
The resulting state, which we shall call photon-subtracted-and-added squeezed state (PSAS),
is obtained by applying the annihilation and creation operators,
${\hat a}^\dagger {\hat a}$, to the squeezed vacuum state
$|\hat{S}(r)\rangle$.
After a straightforward calculation using Eqs.~(\ref{iden}) and the normalization, the PSAS appears to 
be
\begin{equation}
|\psi_{{\hat a}^\dagger {\hat a}}\rangle = {\cal N}_{{\hat a}^\dagger {\hat a}}
 \hat{S}(r)(|0\rangle-\sqrt{2}(\tanh r)^{-1}|2\rangle)\rangle
\label{PSAS}
\end{equation}
where ${\cal N}_{{\hat a}^\dagger {\hat a}}=\{1+2(\tanh r)^{-2}\}^{-1/2}$.
In the same manner, the two-photon subtracted squeezed state (TPSS) can be obtained as
\begin{equation}
|\psi_{{\hat a}^2}\rangle = {\cal N}_{{\hat a}^2} \hat{S}(r)(|0\rangle-\sqrt{2}\tanh 
r|2\rangle)\rangle
\label{TPSS}
\end{equation}
with ${\cal N}_{{\hat a}^2}=\{1+2(\tanh r)^2\}^{-1/2}$.

\subsection{Fidelities against ideal states}

The fidelity, $F=|\langle\psi|\psi_t\rangle|^2$, is a measure of how
close a state $|\psi\rangle$ is to the target state $|\psi_t\rangle$.
It is unity when the two states are identical, while it is zero
when the two are orthogonal to each other.
The fidelity between the TPSS (or PSAS) and the ideal squeezed (or regular) SCS
can be obtained as follows.
 A even SSCS can be expressed as
\begin{equation}
|{\rm SSCS}\rangle = \mathcal{N}_+\hat{S}(r')[|\alpha\rangle
+|-\alpha\rangle],
\end{equation}
where $\quad \mathcal{N}_{+} = [2 + 2e^{-2\alpha^2}]^{-1/2}$.
The fidelity $F$ between the TPSS and the even SSCS can be calculated as
\begin{equation}
\begin{aligned}
F& =| \langle \psi_{a^2}|{\rm SSCS}\rangle|^2 \\
&= 8|\mathcal{N}_{+}\mathcal{N}_{a^2} [ \langle
 0|\hat{S}^{\dag}(r-r')|\alpha\rangle
 + \nu \langle 2|\hat{S}^{\dag}(r-r')|\alpha\rangle]|^2
 \end{aligned}
\end{equation}
where $\nu = -\tanh r$.
This expression can be evaluated with help of the $x$-representation:
\begin{equation}
\langle \psi_1|\psi_2\rangle = \int_{-\infty}^{\infty}
\langle \psi_1|x\rangle \langle x|\psi_2\rangle dx,
\end{equation}
where the relevant wave functions are
\begin{eqnarray}
\langle x| \hat{S}(r-r')|0\rangle &=& (\pi g^2)^{-1/4} e^{-x^2/2g^2}, \nonumber \\
\langle x| \hat{S}(r-r')|2\rangle &=& (4\pi g^2)^{-1/4}
\left(\frac{2x^2}{g^2} -1\right)  e^{-x^2/2g^2}, \nonumber \\
\langle x|\alpha \rangle &=& \pi^{-1/4} e^{-(x-\sqrt{2}\alpha)^2/2},
\end{eqnarray}
where $g = \exp[-(r-r')]$.
After some calculations, we arrive at the fidelity:
\begin{equation}
F_{a^2} =
\frac{8g e^{-\frac{2\alpha^2}{1+g^2}}}{1+g^2}
\Big|\mathcal{N}_{+}\mathcal{N}_{a^2}
\big\{
1 + \nu \frac{1 + 4 \alpha^2 g^2 - g^4}{(1+g^2)^2}\big\}\Big|^2.
\end{equation}
The same approach can be used to derive fidelity
 of states prepared by combined photon subtraction and addition:
\begin{equation}
 F_{a^{\dag}a} =
  \frac{8ge^{-\frac{2\alpha^2}{1+g^2}}}{1+g^2}\Big|
  \mathcal{N}_{+}\mathcal{N}_{a^{\dag}a}
\big\{
1 + \mu \frac{1 + 4 \alpha^2 g^2 - g^4}{(1+g^2)^2}\big\}\Big|^2,
\end{equation}
where
\begin{equation}
 \mathcal{N}_{a^{\dag}a} = [1 + 2 \mu^2]^{-\frac{1}{2}},~~~\mu=\frac{1}{\tanh r}.
\end{equation}

\begin{figure}
\centerline{\psfig{figure=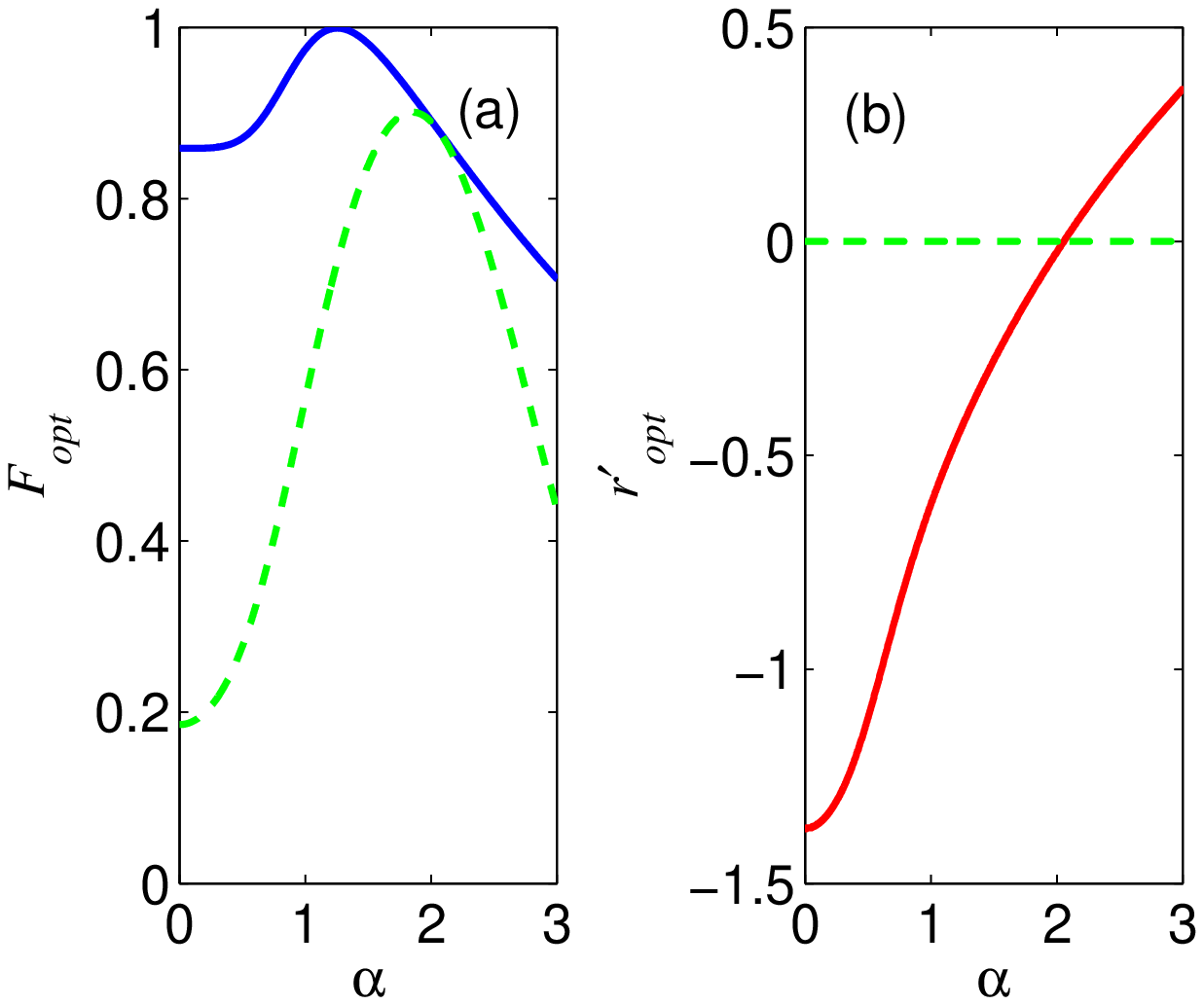,width=9cm}} \caption{(a)
Optimal fidelity $F_{opt}$ between the ideal TPSS ($a^2$) and the squeezed
SCS of amplitude $\alpha$ (solid curve) and the optimal fidelity between the ideal
TPSS and the corresponding regular SCS (dashed curve).
The squeezing
parameter $r^\prime_{opt}$ of the initial squeezed state is $r = -0.7$.
(b) The squeezing parameter $r^\prime$ of the target SSCS for
which the fidelity is optimized. } \label{fidfig1}
\end{figure}
\begin{figure}
\centerline{\psfig{figure=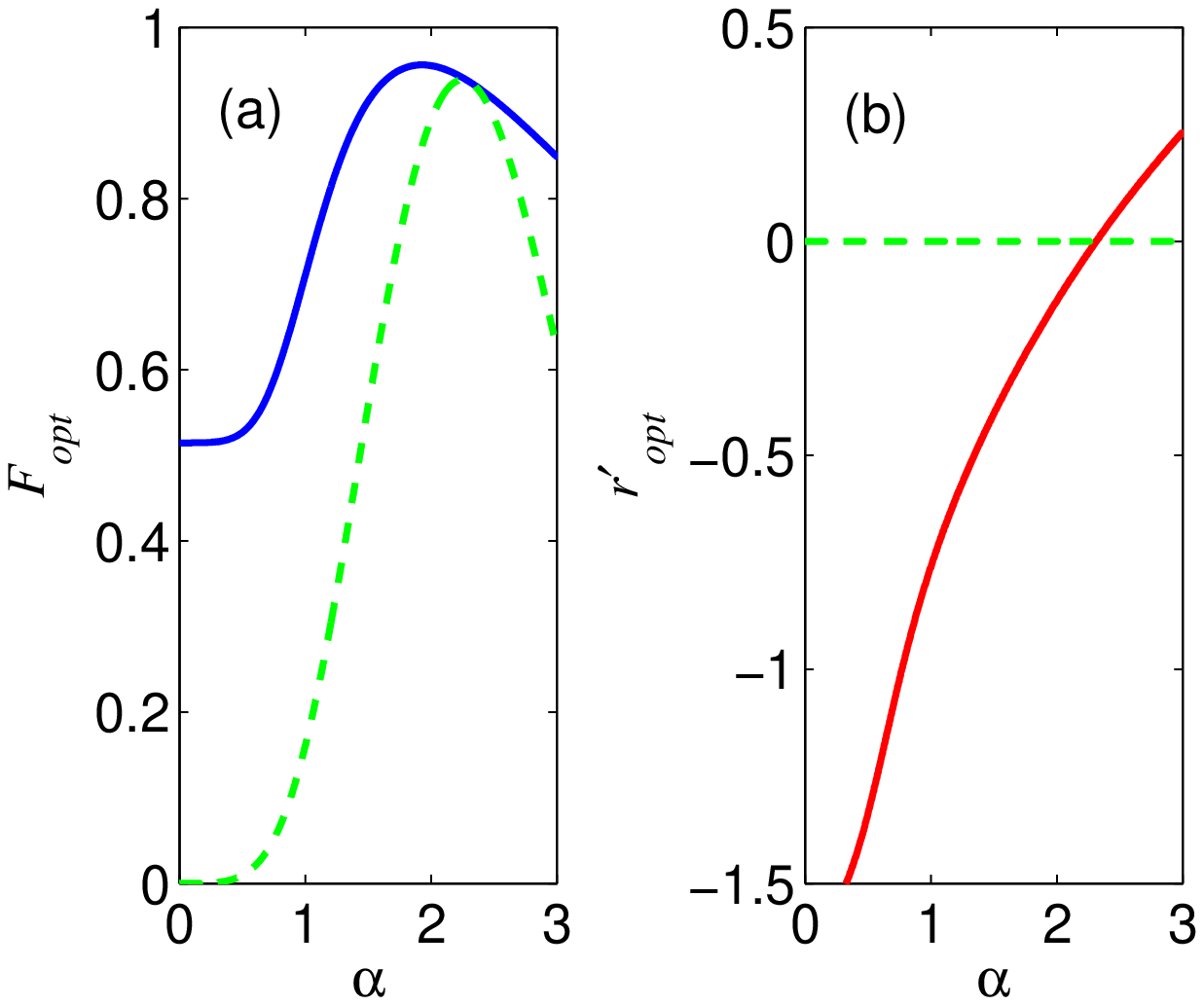,width=9cm}} \caption{(a)
Optimal fidelity $F_{opt}$ between the ideal PSAS ($a^{\dag} a$) and the
SSCS of amplitude $\alpha$ (solid curve) and the optimal fidelity between the ideal PSAS
and the corresponding regular SCS (dashed curve). The squeezing
parameter $r^\prime_{opt}$  of the initial squeezed state is $r = -0.7$. (b) The
squeezing parameter of the target SSCS for which the
fidelity is optimized. }\label{fidfig2}
\end{figure}

In Figs.~\ref{fidfig1} and \ref{fidfig2}, we have used the analytical expressions to show the
optimal fidelities for a range of coherent amplitudes. That is, for each $\alpha$, such
squeezing $r^\prime$ of the SSCS is found to provide the maximal overlap
with the state prepared from initial squeezed state with $r = -0.7$ ($p$-squeezed state,
approximately $6.1$ dB of squeezing, which can be realized using current technology).
For comparison a fidelity with the corresponding regular SCS is also shown.

Figure~\ref{fidfig1}(a) shows the fidelity of the TPSS. It is seen
that although the fidelity tops at $0.9$ for a regular SCS, when a
SSCS is considered, the fidelity of $F = 0.999$ can be
achieved for $\alpha = 1.26$ and $r^\prime = -0.425$.
The squeezing parameter $r^\prime$ of the
target SSCS that optimizes the fidelity against the corresponding
amplitude $\alpha$ is plotted in Fig.~1(b). The fidelity for the
PSAS and the optimizing squeezing parameter against $\alpha$ are
depicted in Fig.~\ref{fidfig2}. The optimal fidelity is $F =
0.956$, which is not as good as the case of the TPSS. However, in
this case, the amplitude of the SSCS is larger as $\alpha=1.93$,
while the optimal squeezing for the SSCS is
only $r^\prime = -0.17$.

\begin{center}
\begin{figure}
\centerline{(a)}
\centerline{\scalebox{0.8}{\includegraphics{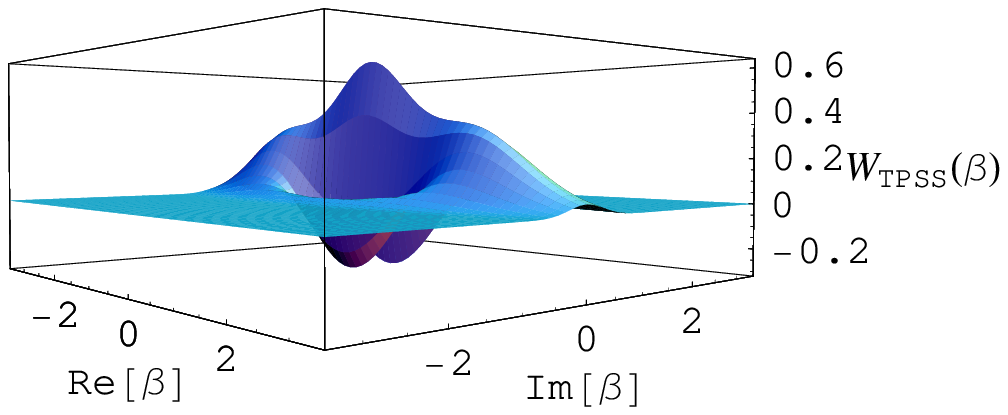}}}
\vspace{0.3cm}
\centerline{(b)}
\centerline{\scalebox{0.79}{\includegraphics{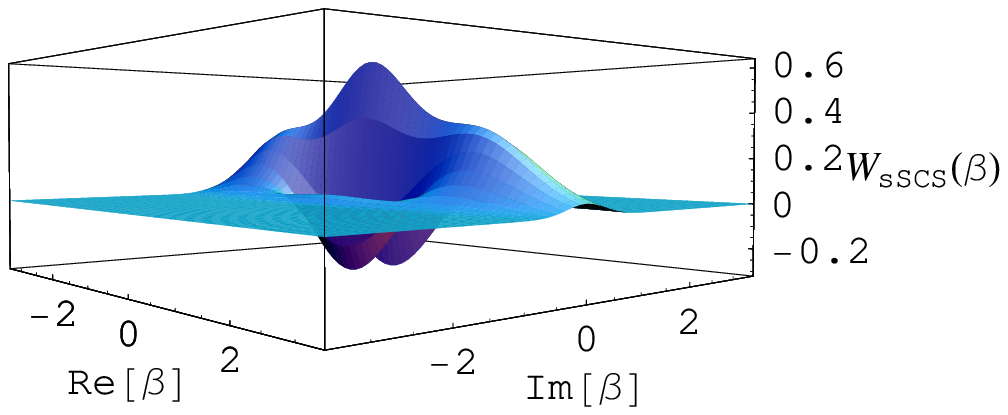}}}
\vspace{0.1cm}
\caption{
(a) The Wigner function of the TPSS with $r=-0.7$ ($\approx$6.1dB)
and (b) the Wigner function of the ideal SSCS
with $\alpha=1.26$ and $r^\prime=-0.425$.
The fidelity between the
two states is nearly perfect as $F> 0.999$.
}
\label{wf1}
\end{figure}
\end{center}

\begin{center}
\begin{figure}
\centerline{(a)}
\centerline{\scalebox{0.8}{\includegraphics{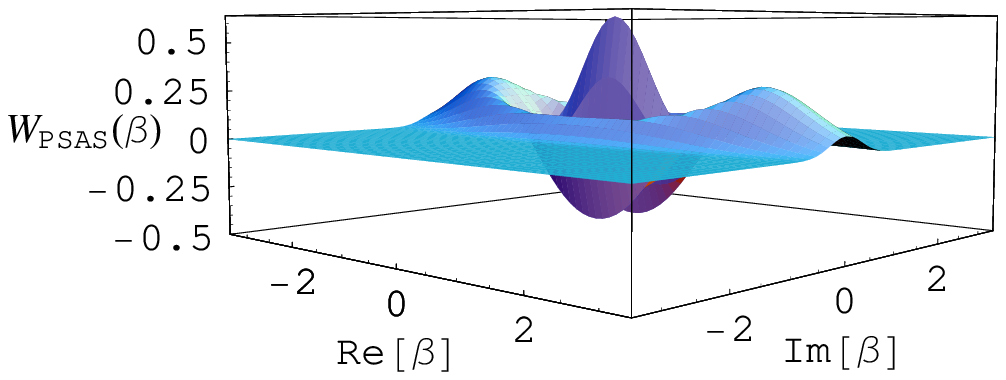}}}
\vspace{0.3cm}
\centerline{(b)}
\centerline{\scalebox{0.79}{\includegraphics{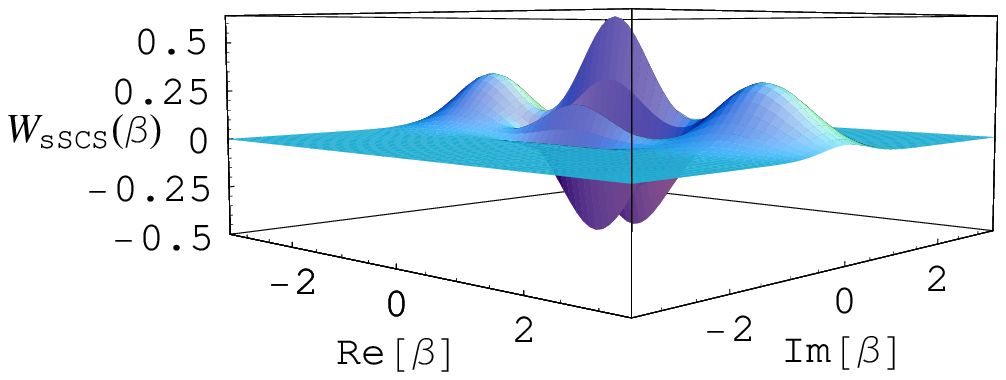}}}
\vspace{0.1cm}
\caption{
(a) The Wigner function of the PSAS (left) with $r=-0.7$ ($\approx$6.1dB)
and (b) the Wigner function of the ideal SSCS
with $\alpha=2$ and $r^\prime=-0.14$ (right).
The fidelity between the two states us $F\approx 0.955$.
}
\label{wf2}
\end{figure}
\end{center}

The Wigner function of a state with density operator $\rho$
can be obtained from the Fourier transform of its
characteristic function $C(\zeta)=\Tr[D(\zeta)\rho]$,
where $D(\zeta)=\exp[\zeta \hat{a}^\dagger-\zeta^* \hat{a}]$ is the displacement operator.
The Wigner functions of the TPSS and PSAS can be obtained
 using Eqs.~(\ref{PSAS}) and (\ref{TPSS}) as
\begin{eqnarray}
  \label{wb1}
&W_{{\hat a}^2}(\beta)={\cal N}_{ss}e^{-2Z}
\Big[1+2\tanh r\big\{Z^\prime\nonumber\\
&~~~~~~~~~~~~~~~~~~~~~~+\tanh r [1+8Z(Z-1)]\big\}\Big],\\
&W_{{\hat a}^\dagger {\hat a}}(\beta)=
{\cal N}_{as}e^{-2Z}\Big[1+2\coth r\big\{Z^\prime\nonumber\\
&~~~~~~~~~~~~~~~~~~~~~~+\coth r [1+8Z(Z-1)]\big\}\Big]
\label{wb2}
\end{eqnarray}
where $Z=e^{2r}\beta_r^2+e^{-2r}\beta_i^2$ and
$Z^\prime=-4e^{2r}\beta_r^2+4e^{-2r}\beta_i^2$. In Figs.~\ref{wf1}
and \ref{wf2}, we consider the squeezing parameter $r=-0.7$
(6.1dB), for the initial squeezed state. Figure~\ref{wf1} shows
again that the TPSS is an extremely good approximation of the
even SSCS. In Fig.~\ref{wf1}, the Wigner functions of the
TPSS of $r=-0.7$ and the even SSCS of $r^\prime=-0.425$
look virtually identical and the fidelity between the two states
is $F>0.999$. Figure~\ref{wf2} presents the Wigner functions of
the PSAS of $r=-0.7$ and the even SSCS of $r^\prime=-0.14$
and $\alpha =2$, where the fidelity between the two states is
$F\approx 0.955$. This shows that slight variation of parameters
still allows for high fidelity.

To conclude, if one is interested in generating SSCSs of high
fidelity, the two-photon subtraction would be a useful
scheme, while the photon subtraction and addition would be a
better strategy to generate large SSCSs.

\section{Experimental considerations}

\subsection{Optical operations with ideal avalanche photodetectors}

So far, we have considered ideal photon addition and subtraction
using annihilation and creation operators. In real experiments,
however, they can be implemented only using approximate schemes.
The setup for implementation of the annihilation (creation) of a
single photon consists of a beam splitter - BS (noncollinear
optical parametric amplification - NOPA) and an avalanche
photodetector. In the Wigner function formalism, the actions of BS
and NOPA, coupling two modes of light, can be characterized with
help of a transformation matrix acting on a vector of variables $
(x_1,p_1,x_2,p_2)$ corresponding to quadrature operators $X_j$ and
$P_j$ with $[X_j,P_{j'}] = i \delta_{jj'}$:
\begin{equation}\label{transm}
 V_{12} = \left(\begin{array}{cccc}
t & 0 & r & 0 \\
0 & t & 0 & \chi \\
-r & 0 & t & 0 \\
0 & -\chi & 0 & t
\end{array}
\right),
\end{equation}
where the generalized parameters are
\begin{eqnarray}
t = \sqrt{T} , \quad r = \chi = \sqrt{1-T}
\end{eqnarray}
for $V_{12}$ to describe the action of a beam splitter with transmissivity $T$, and
\begin{equation}
 t = \sqrt{G}, \quad r = -\chi = \sqrt{G-1}
\end{equation}
for $V_{12}$ to describe a NOPA with amplification gain $G$. The three-mode Wigner
function for the initial squeezed state and two vacuum ancillas is
expressed as
\begin{equation}
W_{\mathrm{tot}}(\xi) =
\frac{1}{(2\pi)^2\sqrt{|\Sigma_{\mathrm{tot}}|}} \exp\left( -
\frac{1}{2} \xi \Sigma_{\mathrm{tot}}^{-1} \xi^T \right).
\end{equation}
Here, the vector of variables for the three-mode Wigner function
is defined as $\xi = (x_1,p_1,x_A,p_A,x_B,p_B)$, where $x=
\beta_r/\sqrt{2}$ and $p= \beta_i/\sqrt{2}$ being compared to the
variable, $z$, used in Eqs.~(\ref{wb1}) and (\ref{wb2}), and the
subscripts of $x$ and $p$ in order to denote the initial and two
ancillary modes, respectively. Furthermore, $|.|$ denotes the
determinant of the matrix, $^T$ stands for transposition and
$\Sigma_{\mathrm{tot}}$ is the covariance matrix of the state,
\begin{equation}
 \Sigma_{\mathrm{tot}} = \diag (V_x,V_p,\frac{1}{2},\frac{1}{2},\frac{1}{2},\frac{1}{2}),
\end{equation}
with $V_x$ and $V_p$ being the variances of the initial squeezed
state.

The initial state subsequently interacts with the two vacuum modes,
transforming the vector of variables into
\begin{equation}
 \xi \rightarrow \xi' = V_B V_A \xi,
\end{equation}
where the transformation matricies, $V_A$ and $V_B$ are of the
form  (\ref{transm}) coupling modes $1$ and $A$ and $1$ and $B$
with parameters $t_A$, $r_A$, $\chi_A$ and $t_B$, $r_B$ and
$\chi_B$, respectivelly. To complete the transformation, a
conditionning measurement is needed. When using two ideal
avalanche photodetectors and post-selecting the state only when
both produce a detection event, we implement a pair of projection
operators $ \hat{1} - |0\rangle\langle 0|$ and the Wigner function
of the output state is transformed to:
\begin{equation}
\begin{aligned}
W_{\mathrm{out}}(x_1,p_1) =& \int W_{\mathrm{tot}}(\xi')[1-2\pi W_{\mathrm{vac}}
(x_A,p_A)]\\
&\times[1-2\pi W_{\mathrm{vac}}(x_B,p_B)]  dx_A dp_A d x_B d p_B
\end{aligned}
\end{equation}
where $W_{\mathrm{vac}}(x,p) = \exp(-x^2-p^2)/\pi$ is the Wigner function of a vacuum state.
This integral can be expressed as a sum of four Gaussian integrals,
\begin{equation}
\label{Wintegral}
\begin{aligned}
W_{\mathrm{out}}(x_1,p_1) = \frac{1}{2\pi\sqrt{|\Sigma_{\mathrm{tot}}|}}&
\Big[I(\Sigma_{nn}) - 2 I(\Sigma_{ny})\\
& - 2 I(\Sigma_{yn}) + 4 I(\Sigma_{yy})\Big],
\end{aligned}
\end{equation}
where $I(\Sigma)$ is a shorthand for
\begin{equation}
 I(\Sigma) = \frac{1}{(2\pi)^2} \int \exp\left( -\frac{1}{2}\xi \Sigma^{-1}
 \xi^T\right) dx_A dp_A dx_B dp_B
\end{equation}
and $\Sigma_{jj'}$ is a covariance matrix for the particular event
when either the two detectors did detect vacuum ($y$) or no
measurement has taken place ($n$). The particular covariance
matrices can be found as
\begin{equation}\label{sigmajj}
 \Sigma_{jj'} = [\Sigma_{\mathrm{tot}}^{-1} + \Pi_{jj'}]^{-1},
\end{equation}
where the $\Pi_{jj'}$ is the semi-inverted covariance matrix of the vacuum state,
\begin{eqnarray}
\Pi_{nn} &= \diag(0,0,0,0,0,0), \quad &\Pi_{yy} = \diag(0,0,2,2,2,2), \nonumber \\
\Pi_{yn} &= \diag(0,0,2,2,0,0), \quad &\Pi_{ny} = \diag(0,0,0,0,2,2). \nonumber \\
\end{eqnarray}
The matrices $\Sigma_{jj'}$ can be decomposed into a block form,
\begin{equation}
\Sigma_{jj'} = \left(
\begin{array}{ll}
A_{jj'} & C_{jj'} \\
C_{jj'}^T & B_{jj'}
\end{array}
\right),
\end{equation}
where the submatrix $A_{jj'}$ corresponds to variables $x_1$,$p_1$, submatrix
$B_{jj'}$ to $x_A,p_A,x_B,p_B$ and submatrix $C_{jj'}$ covers relations between these two groups.
After the integration, we can arrive at the final Wigner function:
\begin{equation}
\label{Wout}
\begin{aligned}
W_{\mathrm{out}}(x,p) =
&\frac{1}{\mathcal{N}\sqrt{|\Sigma_{\mathrm{tot}}|}} \Big[c_{nn}
W_{A_{nn}}(x,p) - 2 c_{ny}  W_{A_{ny}}(x,p)\\& - 2 c_{yn}
W_{A_{yn}}(x,p) + 4c_{yy} W_{A_{yy}}(x,p)\Big],
\end{aligned}
\end{equation}
where
\begin{equation}
W_{\Sigma}(x,p) = \frac{1}{2\pi\sqrt{|\Sigma|}}\exp\left[-\frac{1}{2} (x,p)\Sigma^{-1}(x,p)^T\right]
\end{equation}
is the Wigner function of a Gaussian state with covariance matrix $\Sigma$, the coefficients are
\begin{equation}
c_{jj'} = \sqrt{|\Sigma_{jj'}|}
\end{equation}
and
\begin{equation}
\mathcal{N} = (c_{nn} - 2c_{ny} - 2c_{yn} + 4c_{yy})(\sqrt{|\Sigma_{\mathrm{tot}}|})^{-1}
\end{equation}
is the normalization factor and overall probability of success.

\begin{figure}
\centerline{\psfig{figure=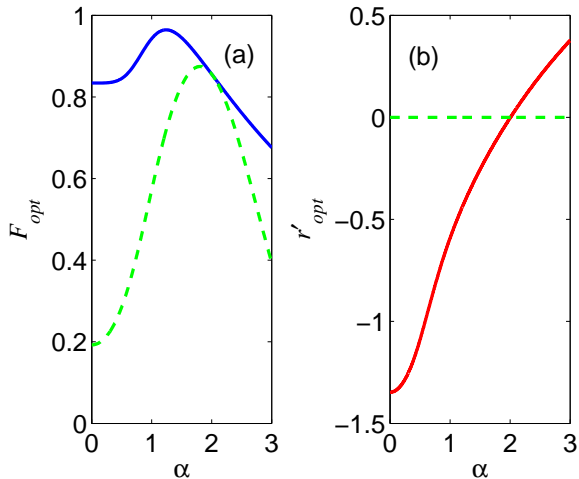,width=9cm}} \caption{(a)
Optimal fidelity $F_{opt}$  between the TPSS ($a^2$) generated using ideal
avalanche photon detectors and the SSCS of amplitude $\alpha$
(solid curve) and the optimal fidelity with the corresponding regular SCS
(dashed curve). The squeezing parameter of the initial squeezed
state is $r = -0.7$. The parameter of real transformation  is
$T=0.99$. (b) The squeezing parameter of the target SSCS
for which the fidelity is optimized. } \label{fidfigB1}
\end{figure}

\begin{figure}
\centerline{\psfig{figure=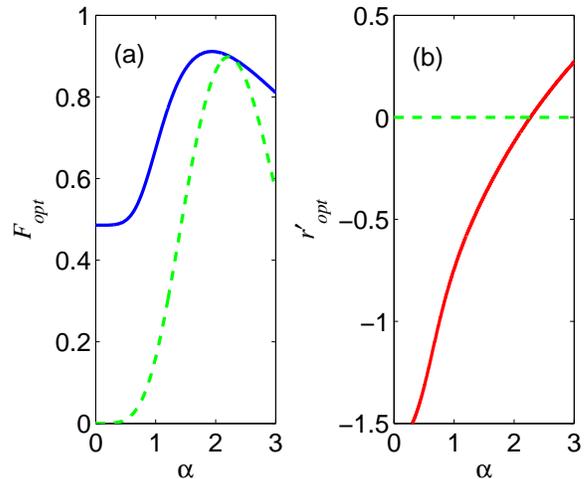,width=9cm}} \caption{Optimal
fidelity $F_{opt}$ between the PSAS ($a^{\dag} a$) generated using ideal
avalanche photon detectors and the SSCS of amplitude $\alpha$ (solid curve)
and the optimal fidelity with the corresponding regular SCS
(dashed curve). The squeezing parameter of the initial squeezed
state is $r = -0.7$. (b) The squeezing parameter of the target
SSCS for which the fidelity is optimized. The parameters
of real transformations are $T=0.99$, $G = 1.01$.
}\label{fidfigB2}
\end{figure}

In order to compare the final state  (\ref{Wout}) with our target state,
the even SSCS, we need to employ its Wigner function
\begin{equation}
\begin{aligned}
&W_{SCS}(x,p) = \mathcal{N}^2_{SCS}\frac{e^{-g p^2}}{\pi}\{
\exp \Big[-\frac{(x-\sqrt{2g}\alpha)^2}{g}\Big]\\
& + \exp \Big[-\frac{(x
+\sqrt{2g}\alpha)^2}{g}\Big] + 2 e^{-x^2/g}\cos(2\sqrt{2g}\alpha p)\},
\end{aligned}
\end{equation}
where $\mathcal{N}_{SCS} = [2+2 e^{-2\alpha^2}]^{-1/2}$,
$\alpha$ is the coherent amplitude, and $g = \exp(-2r)$ characterizes squeezing of the state.
The fidelity can then be calculated as the overlap
between the two Wigner functions:
\begin{equation}
F = 2\pi \int W_{\mathrm{out}}(x,p)W_{\rm SCS}(x,p) dx dp.
\label{wigner-fidelity}
\end{equation}
Since the Wigner function $W_{\rm out}(x,p)$ is a sum of Gaussian functions, we can treat
the integration in parts and express the final fidelity as
\begin{equation}
\label{f1}
\begin{aligned}
F = \frac{1}{\mathcal{N}\sqrt{|\Sigma_{\mathrm{tot}}|}}
&\big[c_{nn} f(A_{nn}) -2 c_{ny} f(A_{ny})\\
&-2 c_{yn}f(A_{yn}) + 4 c_{yy}f(A_{yy})\big].
\end{aligned}
\end{equation}
Here $f(\Sigma)$ denotes the fidelity between a Gaussian state with covariance matrix
$\Sigma$ and a SSCS. Since all the covariance
matrices used are diagonal, we can write these partial fidelities as functions
of the diagonal elements as
\begin{equation}
\label{f2}
\begin{aligned}
f[\diag(\mu,\nu)] = &2 \left[(1+e^{-2\alpha^2})\sqrt{(g+2\mu)(\frac{1}{g}+2\nu) }
\right]^{-1} \\
&\times\left[\exp(\frac{-2\alpha^2}{1+2\mu/g}) + \exp(-\frac{2\alpha^2}{1+1/2g\nu})\right].
\end{aligned}
\end{equation}
With Eqs.~(\ref{f1}) and (\ref{f2}) we can finally obtain the required
fidelity. In analogy with the previous section, the behavior of the
state is depicted on Figs.~\ref{fidfigB1} and \ref{fidfigB2}. The
comparison reveals that these real states behave in a similar
pattern as the ideal ones, the performance is however slightly
worse. For realistic parameters, $T = 0.99$ and $G = 1.01$, the
optimal fidelities are $F \approx 0.96$ for two photon subtraction
and $F \approx  0.91$ for subtraction and addition.


\subsection{Effects of imperfect detectors}
\begin{figure}
\centerline{\psfig{figure=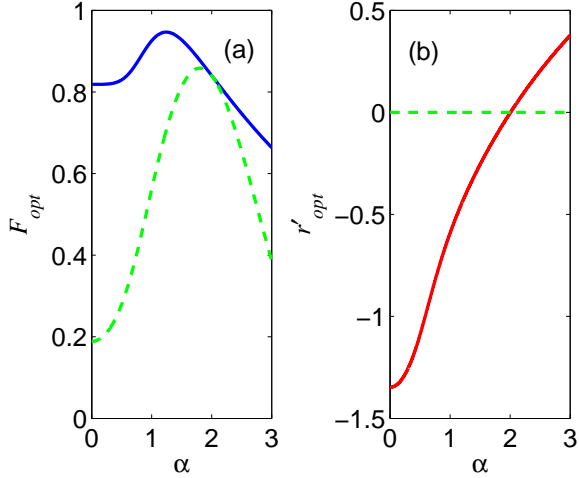,width=9cm}} \caption{(a)
Optimal fidelity $F_{opt}$ between the TPSS ($a^2$) generated using
realistic avalanche photon detectors and the SSCS of amplitude
$\alpha$ (solid curve) and the optimal fidelity with the corresponding regular
SCS (dashed curve). The squeezing parameter of the initial
squeezed state is $r = -0.7$. The parameters of realistic
transformations are $T=0.99$ and $\eta = 0.6$. (b) The squeezing
parameter of the target SSCS for which the fidelity is
optimized.} \label{ff1}
\end{figure}

\begin{figure}
\centerline{\psfig{figure=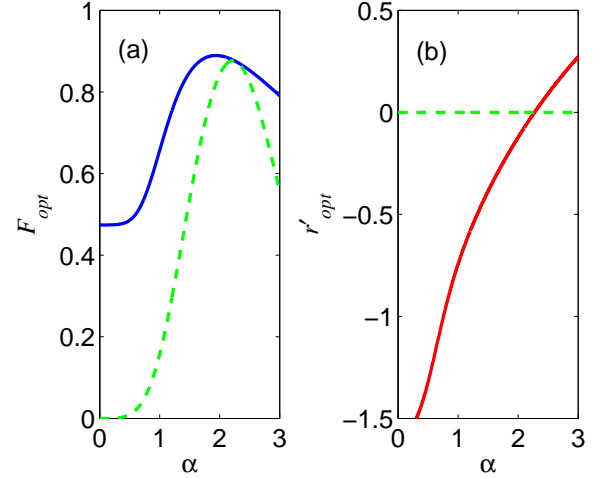,width=9cm}} \caption{(a)
Optimal fidelity $F_{opt}$ between the PSAS ($a^{\dag} a$) generated using
realistic avalanche photon detectors and the SSCS of amplitude $\alpha$
(solid curve) and the optimal fidelity with the corresponding regular
SCS (dashed curve). The squeezing parameter of the initial
squeezed state is $r = -0.7$. (b) The squeezing parameter of the
target SSCS for which the fidelity is optimized. The
parameters of realistic transformations are $T=0.99$, $G = 1.01$
and $\eta = 0.6$.} \label{ff2}
\end{figure}

We have so far considered ideal avalanche photodetectors with unit
quantum efficiency. However, in real experiments, detection
efficiency is always limited. A realistic detector with efficiency
$\eta$ can be modelled by a beam splitter with transmissivity
$\eta$ and vacuum at the idle port inserted in front of an ideal
detector.
The Wigner function of the output state can then be represented as
\begin{equation}\label{imp1}
\begin {aligned}
W_{\mathrm{out}}(x_1,p_1) =
\frac{1}{2\pi\sqrt{|\Sigma'_{\mathrm{tot}}|}}
 &\big[I(\Sigma'_{nn}) - 2 I(\Sigma'_{ny}) \\
 &- 2 I(\Sigma'_{yn}) + 4 I(\Sigma'_{yy})\big]
 \end{aligned}
\end{equation}
where the covariance matrice $\Sigma'_{\mathrm{tot}}$ incorporates
the effect of the imperfect detection as
\begin{eqnarray}\label{imp2}
   \Sigma'_{\mathrm{tot}} &=& \Xi \Sigma_{\mathrm{tot}} \Xi 
    + (1-\eta)
    \mathrm{diag}(0,0,\frac{1}{2},\frac{1}{2},\frac{1}{2},\frac{1}{2}),
    \nonumber \\
    \Xi &=& \mathrm{diag}(1,1,\sqrt{\eta},\sqrt{\eta},\sqrt{\eta},\sqrt{\eta})
\end{eqnarray}
and the matrices $\Sigma'_{j j'}$ can be obtained from
(\ref{imp2}) and (\ref{sigmajj}). Finally, the fidelities can be
arrived at using Eqs.~(\ref{f1}) and (\ref{f2}). Figure~\ref{ff1}
presents the optimal fidelity of the TPSS generated using
avalanche photodetectors with $\eta=0.6$ and a beam splitter of
$T=0.99$. Remarkably, under these realistic assumptions, the
fidelity $F\approx 0.95$ can still be obtained. The fidelity for
PSAS under the same considerations is plotted in Fig.~\ref{ff2}
where the optimal fidelity drops down to $F\approx 0.89$. This
confirms that two-photon subtraction is a better scheme to
generate SSCSs of high fidelity, even though the amplitudes
are smaller in this case.

\section{successive applications of photon subtraction}

\begin{figure}
\centerline{\scalebox{0.82}{\includegraphics{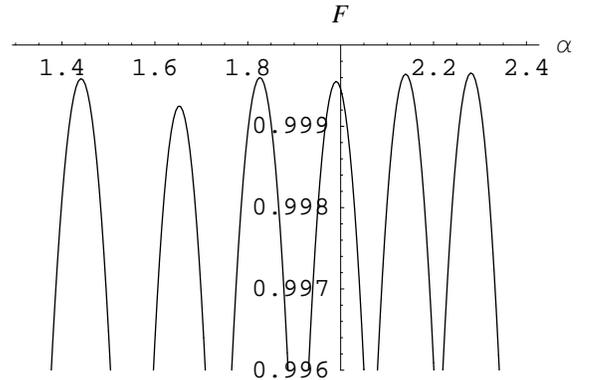}}}
\vspace{0.1cm}
\caption{The optimal fidelity between the NPSSs with ${\cal R}(=T\tanh r)=0.6$
and the SSCSs of $r^\prime=0.44$
from $N=3$ (left) to $N=8$ (right). The $x$ axis represents the amplitude $\alpha$
of the target SSCS. The optimal fidelity is $F>0.999$ regardless of the number of $N$. }
\label{fig:fn}
\end{figure}

We have shown that the SSCS is extremely well
approximated by the TPSS.
We now show that this can be generalized to arbitrary
$N$-photon subtraction with $N\geq 3$.
Namely, $N$-photon-subtracted squeezed states (NPSSs) are
good approximations of the SSCSs,
which may be compared with the proposal by Fiur\'a\v{s}ek {\it et al.}
to generate an arbitrary state by photon subtractions and displacements of
a squeezed state \cite{FK}.
We suppose that a beam splitter of transmmittivity $T$
and an ideal photodetector is used to subtract $N$ photons from a squeezed state.
The Wigner function of the NPSS is then obtained as \cite{Dakna97}
\begin{equation}
\begin{aligned}
W_N(\beta)=&{\cal M}_N\exp\Big(-\lambda\beta_r^2-\frac{\beta_i^2}{\lambda}\Big)
\sum_{k=0}^N\frac{(-2|{\cal R}|)^k}{k![(N-k)!]^2}\\
&\times\Big|H_{N-k}[i\sqrt{{\cal R}\lambda}(\beta_r+i\frac{\beta_i}{\lambda})]\Big|^2
\end{aligned}
\end{equation}
where
\begin{eqnarray}
&&{\cal M}_N=\sum_{k=0}^{N/2}\frac{(2|{\cal R}|)^{N-2k}}{(N-2k)!(k!)^2},\\
&&\lambda=(1-{\cal R})/(1+{\cal R}),\\
&&{\cal R}=T |\tanh r|^2
\end{eqnarray}
and $H_n[x]$ is the Hermite polynomial.
When $N$ is odd (even), the NPSS should be compared with the odd (even) SCS.
The fidelity between the NPSS and the SSCS can be obtained using Eq.~(\ref{wigner-fidelity}).
It is nontrivial to obtain analytical expressions of the fidelity
for an arbitrary $N$. As presented in Fig.~\ref{fig:fn},
we numerically assess the fidelity $F$ and plot
it from $N=3$ to $N=8$ against the amplitude $\alpha$ of the SSCS.
We suppose $\cal R$ is 0.6, i.e. $r\approx 0.7$ (6.1dB)
when $T=0.99$ and $r=0.75$ (6.5dB) when $T=0.95$.
The degrees of squeezing, 6.1dB to 6.5dB, are experimentally achievable.
Interestingly, the squeezing parameter of the SSCS is $r^\prime=0.44$
regardless of $N$.
Our numerical calculation confirms this generalization up to $N=15$.
For example, the fidelity is as high as $F>0.999$ for $\alpha=2.52$ ($\alpha=3$)
when $N=10$ ($N=15$) with the same squeezing parameter
$r^\prime=0.44$ of the SSCS.

\section{Remarks}

We have pointed out the twofold directions of the development for
the generation of SCSs. The generation of SCSs with high
fidelities ($F>0.99$) and those with large amplitudes
($\alpha\gtrsim 2$) aims at practical applications of quantum
information processing and macroscopic tests of quantum theory.
The SSCSs, which were generated in recent experiments \cite{Cat07},
may be a good alternative to the SCSs for the aforementioned purposes while their
 fidelity in those experiments is yet to be improved.
In this paper, we have studied how photon addition and subtraction
can be used to generate the SSCSs with high fidelities and large
amplitudes.
We have found that the single photon subtraction and
subsequent addition with a squeezed vacuum state can cause the
production of the approximate SSCS with $F\approx 0.956$
and $\alpha\approx2$ when the squeezing degree is about 6.1dB
which is achievable using current technology. Furthermore, we show
that $N$ photon subtraction may be used to generate the SSCSs
with extremely high fidelities as $F>0.999$. The amplitude of
the SSCS increases as the number of the subtracted photons
gets larger. For example, $\alpha=1.26$ is obtained for two-photon
subtraction while $\alpha=1.68$ for four-photon subtraction when
the fidelity is $F>0.999$ for both cases.

We have assessed some experimental imperfections in implementing
photon addition and subtraction such as inefficiency of
photodetectors and nonunit transmmittivity of beam splitters. It
has been shown that the fidelity of $F\approx 0.95$ can still be
obtained with detection efficiency $\eta=0.6$ and beam-splitter
trasmmittivity $T=0.99$ for two-photon subtraction to generate
SSCSs. We believe that our work will immediately motivate
experimental efforts to generate high-fidelity SCSs and
large-amplitude ones that are useful for various quantum
information applications and fundamental studies.

\acknowledgements

This work was supported by the UK EPSRC and QIPIRC and
by the Korea Science and Engineering Foundation 
(KOSEF) grant funded by the Korea government(MEST) (R11-2008-095-01000-0).
P.M. acknowledges support from the European Social Fund and from the Ministry 
of Education of the Czech Republic (Grants No. LC06007 and No. MSM6198959213). 
He also acknowledges support of the Future and Emerging Technologies (FET) 
programme within the Seventh Framework Programme for
Research of the European Commission, under the FET-Open grant 
agreement COMPAS, number 212008.


\begin{thebibliography}{99}


\bibitem{Yurke} B. Yurke and D. Stoler, \prl {\bf 57}, 13 (1986).

\bibitem{WScat} W. Schleich, M. Pernigo, and F.L. Kien,
Phys. Rev. A {\bf 44}, 2172 (1991).

\bibitem{Schr}
E. Schr\"odinger, {\it Naturwissenschaften.} {\bf 23},  807-812; 823-828; 844-849 (1935).

\bibitem{Derek} D. Wilson, H. Jeong, and M. S. Kim, J. Mod
  Opt {\bf 49} 851 (2002).

\bibitem{jeongsonkim} H. Jeong,
W. Son, M. S. Kim, D. Ahn, and C. Brukner,
Phys. Rev. A {\bf 67}, 012106 (2003).

\bibitem{Magda}
M. Stobi\'nska, H. Jeong, T. C. Ralph,
Phys. Rev. A {\bf 75}, 052105 (2007).

\bibitem{Enk01}
 S. J. van Enk and O. Hirota, Phys. Rev. A. {\bf 64},
  022313 (2001).

\bibitem{JKL01}
H. Jeong, M. S. Kim, and J. Lee, Phys. Rev. A. {\bf
64}, 052308 (2001).

\bibitem{puri} H. Jeong and M. S. Kim,
Quantum Information and Computation {\bf 2}, 208 (2002).

\bibitem{Jeong02}
H. Jeong and M. S. Kim
Phys. Rev. A 65, 042305 (2002).

\bibitem{Ralph03}
T. C. Ralph, A. Gilchrist, G. J. Milburn, W. J. Munro, and S. Glancy,
Phys. Rev. A  {\bf 68}, 042319 (2003).

\bibitem{WeakForce}
W. J. Munro, K. Nemoto, G. J. Milburn, and S. L. Braunstein,
 Phys. Rev. A {\bf 66}, 023819 (2002).

\bibitem{gerry-added} C. C. Gerry, Phys. Rev. A 59, 4095 (1999).
 
\bibitem{Dakna} M. Dakna, T. Anhut, T. Opatrn\'y, L. Kn\"oll, and
D.-G. Welsch,   Phys. Rev. A. {\bf  55}, 3184 (1997).

\bibitem{Dakna2} M. Dakna, J. Clausen, L. Kn\"oll, and D.-G.
Welsch, \pra {\bf 59}, 1658 (1999).

\bibitem{Lund04} A. P. Lund, H. Jeong, T. C. Ralph, and M. S. Kim, \pra
  {\bf 70}, 020101(R) (2004).

\bibitem{jc1}  H. Jeong, M.S. Kim, T.C. Ralph, and B.S. Ham,
Phys. Rev. A {\bf 70}, 061801(R) (2004) .

\bibitem{jc2}  H. Jeong, Phys. Rev. A {\bf 72}, 034305 (2005).

\bibitem{jc3} A. M. Lance, H. Jeong, N. B. Grosse, T. Symul, T. C. Ralph, and P. K. Lam,
Phys. Rev. A {\bf 73}, 041801(R) (2006).

\bibitem{jc4} H. Jeong, A. M. Lance, N. B. Grosse, T. Symul, P. K. Lam, and T. C. Ralph,
Phys. Rev. A {\bf 74}, 033813 (2006).

\bibitem{KM} A. E. B. Nielsen and K. M{\o}lmer,
quant-ph/arXiv:0708.1956.

\bibitem{JLR05}
H. Jeong, A. P. Lund, and T. C. Ralph,
Phys. Rev. A {\bf 72}, 013801 (2005)


\bibitem{kt1} J. Wenger, R. Tualle-Brouri, and P. Grangier, \prl
      {\bf 92}, 153601 (2004).

\bibitem{kt2}
A. Ourjoumtsev, R. Tualle-Brouri, J. Laurat, and Ph. Grangier,
{\bf 312} 83 (2006).

\bibitem{kt3} J. S. Neergaard-Nielsen, B. M. Nielsen, C. Hettich,
K. M{\o}lmer, and E. S. Polzik,
Phys. Rev. Lett. {\bf 97}, 083604 (2006).

\bibitem{kt4} K. Wakui, H. Takahashi, A. Furusawa, and M. Sasaki
Optics Express {\bf 15}, 3568 (2007).

\bibitem{Kim05} M. S. Kim, E. Park, P. L. Knight, and H. Jeong,
Phys. Rev. A {\bf 71}, 043805 (2005).

\bibitem{OP} S. Olivares and M. G. A. Paris,
J. Opt. B: Quantum Semiclass. Opt. {\bf 7} S616 (2005);
Laser Physics, {\bf 16}, 1533 (2006).

\bibitem{SS06}
S. Suzukia, K. Tsujinoa, F. Kannarib, and M. Sasaki,
Optics Communications
{\bf 259} 758 (2006).


\bibitem{Cat07}
A. Ourjoumtsev, H. Jeong, R. Tualle-Brouri, and Ph. Grangier,
Nature {\bf 448}, 784 (2007).

\bibitem{suitability}
P. Marek and M. S. Kim, Phys. Rev. A {\bf 78}, 022309 (2008).

\bibitem{Sasaki1}
M. Sasaki, M. Takeoka, and H. Takahashi, Phys. Rev. A 77, 063840 (2008);
M. Takeoka, H. Takahashi, and M. Sasaki,
Phys. Rev. A 77, 062315 (2008).

\bibitem{Sasaki2}
H. Takahashi, K. Wakui, S. Suzuki, M. Takeoka, K. Hayasaka, A. Furusawa, M. Sasaki,
arXiv:0806.2965.

\bibitem{Lund07} A. P. Lund, T. C. Ralph, H. L. Haselgrove,
Phys. Rev. Lett. 100, 030503 (2008).

\bibitem{Serafini}
A. Serafini, S. De Siena, F. Illuminati, and
M. G. A. Paris, J. Opt. B: Quantum Semiclass. Opt. {\bf 6}, S591 (2004).

\bibitem{squeez1}
R. Filip, P. Marek, and U. L. Andersen, \pra {\bf 71}, 042308
(2005); J. I. Yoshikawa, T. Hayashi, T. Akiyama, N. Takei, A. Huck,
U. L. Andersen, and A. Furusawa, Phys. Rev. A {\bf 76}, 060301(R)
(2007).

\bibitem{BRbook} S. M. Barnett and P. M. Radmore,
{\it Methods in Theoretical Quantum Optics},
Oxford University Press (1997).


\bibitem{Dakna97} M. Dakna, T. Anhut, T. Opatrn\'{y}, L. Kn\"{o}ll, and D.-G. Welsch,
 Phys. Rev. A 55, 3184 - 3194 (1997)

\bibitem{FK} J. Fiur\'a\v{s}ek, R. Garc\'ia-Patr\'on, and N. J. Cerf,
Phys. Rev. A {\bf 72}, 033822 (2005); M. S. Kim, J. Phys. B {\bf 41},
133001 (2008).

\end{thebibliography}
\end{document}